\documentclass{aa}
\usepackage{txfonts} \usepackage{graphicx}
\usepackage{color}

\begin{document}


\title{Nonradial and radial period changes of the Delta Scuti star 4~CVn\\ II. Systematic behavior over 40 years}
 
\author{M.~Breger\inst{1,2} \and M. H. Montgomery \inst{2} \and P.~Lenz\inst{1,3}  \and A.~A.~Pamyatnykh\inst{3}}

\institute{Institut f\"ur Astrophysik der Universit\"at Wien, T\"urkenschanzstr. 17, A--1180 Wien, Austria\\
        \email{breger@astro.as.utexas.edu}
\and
Department of Astronomy, University of Texas, Austin, TX 78712, USA
\and
N. Copernicus Astronomical Center, Bartycka 18, 00-716 Warsaw, Poland}

\date{Received date; accepted date}

\abstract 
{}
{Radial and nonradial pulsators on and near the main sequence show period and amplitude changes that are too large to be the product of stellar evolution.
The multiperiodic Delta Sct stars are well suited to study this, as the period changes of different modes excited in the same star can be compared.
 This requires a very large amount of photometric data covering years and decades as well as mode identifications.}
{We have examined over 800 nights of high-precision photometry of the multiperiodic pulsator 4~CVn obtained from 1966 through 2012. Because most of the data were obtained in adjacent observing seasons, it is possible to derive very accurate period values for a number of the excited pulsation modes and to study their systematic changes from 1974 to 2012.}
{Most pulsation modes show systematic significant period and amplitude changes on a timescale of decades. For the well-studied modes, around 1986 a general reversal of the directions
of both the positive and negative period changes occurred. Furthermore, the period changes between the different modes are strongly correlated, although they differ in size and sign. 
For the modes with known values of the spherical degree and azimuthal order, we find a correlation between the direction of the period changes and the identified azimuthal order, $m$. 
The associated amplitude changes generally have similar timescales of years or decades, but show little systematic or correlated behavior from mode to mode.}
{A natural explanation for the opposite behavior of the prograde and retrograde modes is that their period changes are driven by a changing rotation profile. The changes in the rotation profile could in turn be driven by processes, perhaps the pulsations themselves, that redistribute angular momentum within the star. In general, different modes have different rotation kernels, so this will produce period shifts of varying magnitude for different modes.}


\maketitle

\section{Introduction}

Period and amplitude changes are common among the pulsating $\delta$ Sct stars. These changes are not caused by stellar evolution, for example, Breger \& Pamyatnykh (1998) showed that for Pop. I radial pulsators, the measured (1/P) dP/dt values are around 10$^{-7}$ year$^{-1}$ with equal distribution between period increases and decreases. The evolutionary models, on the other hand, predict that the vast majority should show increasing periods with values a factor of ten smaller than observed. For nonradial $\delta$ Sct pulsators of Pop. I, the discrepancies are even larger. The behavior suggests that for these relatively unevolved stars the rate of evolution cannot be deduced from the period changes. Since that study, additional attempts to match the observed changes to evolutionary models have also failed. A recent example is that of Murphy et al. (2012), who attempt to explain the observed amplitude changes in the star KIC 3429637 ($\rho$ Puppis subclass) by evolutionary changes in the pulsation driving.

It is, therefore, essential to study the period and amplitude changes in detail, in order to separate the evolutionary from the other effects. These `other effects' are at present poorly understood and presumably temporary changes in the structure of the star. Such investigations  require extensive data covering decades. We refer to promising results for the cepheid Polaris, where the long-term changes of 4.5s yr$^{-1}$, covering 150 years, are evolutionary, while smaller changes are superposed, for example, around 1965 (Turner et al. 2005).

In $\delta$ Sct stars, such studies are made more difficult by the presence of hundreds of simultaneously excited pulsation modes of generally small amplitudes. Even when the studies concentrate on the large-amplitude modes only, these stars often show close frequencies from modes of different spherical degree, $\ell$, caused by mode trapping in the envelope (Breger, Lenz \& Pamyatnykh 2009). However, the presence of radial as well as nonradial modes can also be advantageous to distinguish between different astrophysical explanations. Pulsation modes of different degree probe different parts of the star and their period changes show changes in the stellar structure. Changes in stellar radius and/or rotation lead to changes in the observed rotational splitting and thereby a dependence of the period changes on the azimuthal order, $m$, of a pulsation mode. Furthermore, magnetic cycles also lead to period changes.

One of the most promising and best-studied stars is the evolved $\delta$ Sct variable 4~CVn (HR~4715 = HD 107904 = AI CVn, F3III-IV). Its variability was discovered by Jones \& Haslam (1966). This led to a large number of photometric studies, of which some were extensive multisite campaigns. Most of the early photometry obtained prior to 2000 have been summarized in Table 1 of Breger (2000). Since then, additional photometry from 1991 was made available by R. J. Dukes (Breger, Davis \& Dukes 2008).

The largest photometric study of 4 CVn was obtained in two different passbands using a dedicated automatic photoelectric telescope. It covered 702 nights from 2005 through 2012 (Breger 2016, hereafter called Paper I). 64 pulsation frequencies could be detected at a high significance level. During these eight years, the periods of the well-studied dominant nonradial modes were found to be variable with rates of change
ranging from (1/P)dP/dt $= -16\times 10^{-6}$ to $+13\times 10^{-6}$ year$^{-1}$. These values are one or two orders of magnitude larger than expected from stellar evolution.
Furthermore, both increasing and decreasing periods were found. However, for each of the more dominant modes studied, the rates of period changes were constant from month to month as well as from year to year.

The question arises as to whether these rates of period change are constant over longer time scales than the previously studied eight years. Due to the large sizes of the observed period changes, we would expect sign reversals at some point. Do such reversals occur simultaneously in all pulsation modes, irrespective of the sign of the period changes? Such a situation would suggest a change in the internal structure of the star as the origin of the period changes.

The available photometric data for 4 CVn cover four decades. In this paper we examine the period changes of the main pulsation modes of 4 CVn over these decades and correlates the result with the type of pulsation mode.

\section{Photometric frequency determinations}

In this investigation we have examined systematic frequency changes on the
order of 10$^{-5}$ year$^{-1}$ or smaller. This means that the data covering many
decades need to subdivided into a number of separate time groups,
each of which is suitable for precise multifrequency solutions. Each group needs
to contain a large number of high-precision measurements to yield accurate multifrequency
solutions and span a sufficiently long time-base for high frequency resolution.
These requirements suggest that each group should contain two successive observing seasons, if possible.

\begin{table}
\caption{Photometric observations used}
\begin{tabular}{lcll}
\hline
\noalign{\smallskip}
Year & Nights used & Observer/reference & Notes\\
\noalign{\smallskip}
\hline
\noalign{\smallskip}
1974     &      24      &       Fitch (1980)    &       Note 1  \\
1976     &      4       &       Loumos (1980)   &       Note 2  \\
1976     &      7       &       Fitch (1980)    &       Note 1  \\
1977 &  9       &       Loumos (1980)   &       Note 2  \\
1977     &      5       &       Warman et al. (1979)    &               \\
1978     &      6       &       Loumos (1980)   &       Note 2  \\
1983     &      8       &       Breger et al. (1990)    & Multisite             \\
1984     &      14      &       Breger et al. (1990)    & Multisite     \\
1991     &      32      &       Breger, Davis \& Dukes (2008)   \\
1992     &      6       &       Davis \& Duke   & Note 3                \\
1996     &      53      &       Breger et al. (1999)    &       Multisite       \\
1997 &  32      &       Breger \& Hiesberger (1999)     &               \\
2005     &      62      &       Breger (2016)   &               \\
2006     &      63      &       Breger (2016)   &       \\
2007     &      86      &       Breger (2016)   & Note 4                \\
2008     &      87      &       Breger (2016)   & Note 4                \\
2009     &      76      &       Breger (2016)   & Note 4                \\
2010     &      96      &       Breger (2016)   & Note 4                \\
2011     &      108     &       Breger (2016)   & Note 4                \\
2012 &  124     &       Breger (2016)   & Note 4                \\
\noalign{\smallskip}
\hline
\noalign{\smallskip} \end{tabular}
\newline
\begin{flushleft}

1. The unpublished, extensive data obtained by Fitch in 1974 were slightly edited with several
nights with large scatter removed. We also pre-whitened the low frequencies of 1.321 and 1.404~d$^{-1}$,
since these frequencies originate in one of the comparison stars, HD~108100 (Breger et al. 1997). \\
2. In the 1976 -- 1978 analyses, the Loumos data were given double weight
due to their high precision relative to the other data sets in this time period.\\
3. 1992: six nights of unpublished data from a continuation of the 1991 campaign added.\\
4. The observing seasons of 4 CVn start before January 1, for example, the observing season called '2012' covers
measurements obtained during 124 nights from 2011 December to 2012 June.\\

\end{flushleft}
\end{table}
We are fortunate that more than 800 nights of observation are available and that they
naturally divide into several groups. However, a small amount of data could not be used for
the present analysis. Earlier photometry covering 1966 - 1970 (see Table 1 in Breger 2000)
was not extensive enough to determine the frequencies of 4 CVn with sufficient accuracy.
Furthermore, some additional data from 1997 (Stankov et al. 2000) were not included because
of relatively lower internal consistency. We note though that inclusion of these data would
not change the conclusions of the present paper.

Table 1 lists the photometric data used in the present analysis. The pulsation frequency determinations were performed with a package of computer
programs (PERIOD04, Lenz \& Breger 2005). PERIOD04 utilizes both Fourier and multiple-least-squares algorithms,
which do not rely on sequential prewhitening. It is especially suited for extensive data containing large numbers of simultaneously excited frequencies.

In order to examine the period changes in stars, the standard (O-C) phase-shift technique is usually applied. As part of this
technique, an optimum frequency is determined. This frequency is then used to derive the differences between
the actually observed and calculated phases as a function of time.
However, for 4 CVn the (O-C) plots revealed the common difficulty with the phase shift method: the unknown number of elapsed cycles
between the different groups of years with sufficient data. While the photometric coverage of 4 CVn is excellent for
a large number of individual groups of years, the data also show large gaps without coverage between these groups of years.

Consequently, for the present analysis we have not used the (O-C) phase-shift method, but actually determine the optimum values of the main frequencies for each individual data set. This stand-alone approach also
eliminates the potential wavelength-dependent phase-shift problem of the extensive 1974 data, which were obtained through the Str\"omgren $b$ bandpass, as opposed to the Johnson $V$ or Str\"omgren $y$ bandpasses used for the other data sets.

\begin{table}
\caption{Phase differences, amplitude ratios and mode identifications}
\begin{tabular}{lccccc}
\hline
\noalign{\smallskip}    
\multicolumn{2}{c}{Frequency} & Phase difference & Amplitude ratio& $\ell$ & $m$ \\
& cycle~d$^{-1}$ &$\phi_v-\phi_y$ [$^{\rm o}$]  & $v/y$\\

\noalign{\smallskip}
\hline
\noalign{\smallskip}
f$_{1}$ &       8.594   &       -2.72   $\pm$   0.11    &       1.469   $\pm$ 0.002   &       1       &       1       \\
f$_{2}$ &       5.048   &       -2.01   $\pm$   0.10    &       1.543   $\pm$ 0.002   &       1       &       -1      \\
f$_{3}$ &       5.850   &       -6.84   $\pm$   0.12    &       1.460   $\pm$ 0.002   &       2       &       1       \\
f$_{4}$ &       5.532   &       -4.71   $\pm$   0.29    &       1.500   $\pm$ 0.005   &       2       &       1, 2    \\
f$_{5}$ &       7.376   &       -2.87   $\pm$   0.22    &       1.514   $\pm$ 0.006   &       1       &       -1      \\
f$_{6}$ &       6.976   &       0.76    $\pm$   0.30    &       1.540   $\pm$ 0.005   &       0       &       0       \\
f$_{7}$ &       7.552   &       -3.27   $\pm$   0.25    &       1.467   $\pm$ 0.005   &       1       &               \\
f$_{8}$ &       6.680   &       -3.31   $\pm$   0.32    &       1.438   $\pm$ 0.007   &       1       &               \\
f$_{9}$ &       6.117   &       -5.27   $\pm$   0.17    &       1.545   $\pm$ 0.005   &       2       &               \\
f$_{10}$        &       6.191   &       -2.86   $\pm$   0.30    &       1.561   $\pm$ 0.009   &       1       &               \\
f$_{11}$        &       6.440   &       -2.66   $\pm$   0.30    &       1.522   $\pm$ 0.006   &       1       &               \\
\noalign{\smallskip}
\hline
\end{tabular}
\end{table}

\begin{figure}[ht]
\includegraphics[width=\columnwidth]{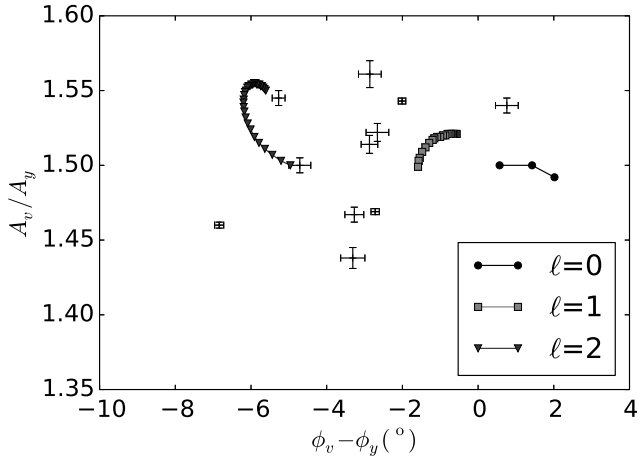}
\caption{Amplitude ratios and phase differences in the Str\"omgren $v$ and $y$ passbands for different pulsation modes. The error bars are determined by comparing results from eight different years.
In A/F stars the observed phase differences are strong indicators of the spherical degree. The results from a theoretical model (see text) covering the frequency range from 4.5 to 9.5 cycle~d$^{-1}$ are also shown.}
\label{Fourier}
\end{figure}

The quantity and spacing of these data prior to the year 2000 immediately suggest the
following five groups: the 1974 data, combined 1976-1978 data,  multisite 1983 and 1984 campaigns,
the 1991 and 1992 data, and the 1996 multisite campaign combined with the extensive 1997 single-site data.
For the 1976-1978 solution, one frequency value was unreliable and had to be omitted:
during 1976, the amplitude of the 7.376 cycle~d$^{-1}$ mode decreased to become essentially undetectable before it started to increase its amplitude
again. For the 2005-2012 seasons we have adopted the following approach to obtain the desired frequency resolution:
for each solution we combined two adjacent observing seasons, covering approximately
18 months. The eight observing seasons from 2005 to 2012 therefore lead to seven solutions.

To determine the period changes, we have calculated new multifrequency fits for these data sets. Additional details on the multiple data sets and their properties can be found in the references listed in Table 1.

Schmid et al. (2014) also discovered that 4 CVn is an eccentric binary system with an orbital period of 124.4 $\pm$ 0.03 d. This raises the question whether the measured pulsation periods and period changes are seriously affected by the light-time effects originating in the binary orbit. Consequently, we repeated our analyses without the orbital corrections. No significant changes in the derived frequencies and amplitudes presented in this paper were found, although the residuals of the analysis without the orbital light-time corrections were slightly higher. Also, the orbital period can be deduced from an analysis of the small
residuals of the uncorrected photometry. This remarkable effect is examined in more detail in a later section. We emphasize that ignoring the binary nature would not change the results and conclusions of the present paper. The numbers presented in this paper are based on the data after applying orbital corrections.

\section{Pulsation mode identifications for 4 CVn}

In order to understand observed intrinsic period changes, it is important to identify the individual observed pulsation modes in terms of the pulsational degree, $\ell$, and the azimuthal order, $m$. For $\delta$ Sct stars, spectroscopic and photometric methods complement each other for mode identification.
In a pioneering study, Schmid et al. (2014) analyzed the line-profile variations of 4 CVn from over 2000 high-dispersion spectra obtained from 2008 to 2011. This study was an extension of an
earlier spectroscopic project by Castanheira et al. (2008). For seven pulsation modes, Schmid et al. determined the $m$ values.

For A and F stars, the brightness variations measured through two different bandpasses show small phase shifts and different amplitudes, depending on the type of pulsation mode. These phase shifts and amplitude ratios are mainly a function of the spherical degree, $\ell$. Consequently, the spectroscopic and photometric mode identification techniques can be combined to determine the spherical degree and azimuthal order for the pulsation modes.

The 702 nights of high-accuracy photometry, obtained from 2005 to 2012 (see Paper I), were obtained with two Str\"omgren $v, y$ filters.
The bandpasses of these filters are ideal to determine the phase shifts and amplitude ratios of stars of spectral type A and F.
In order to minimize any effects due to amplitude and frequency changes, we have treated the data from each annual observing season separately.

\begin{figure*}
\includegraphics[bb = 0 20 630 710,width=\textwidth, clip]{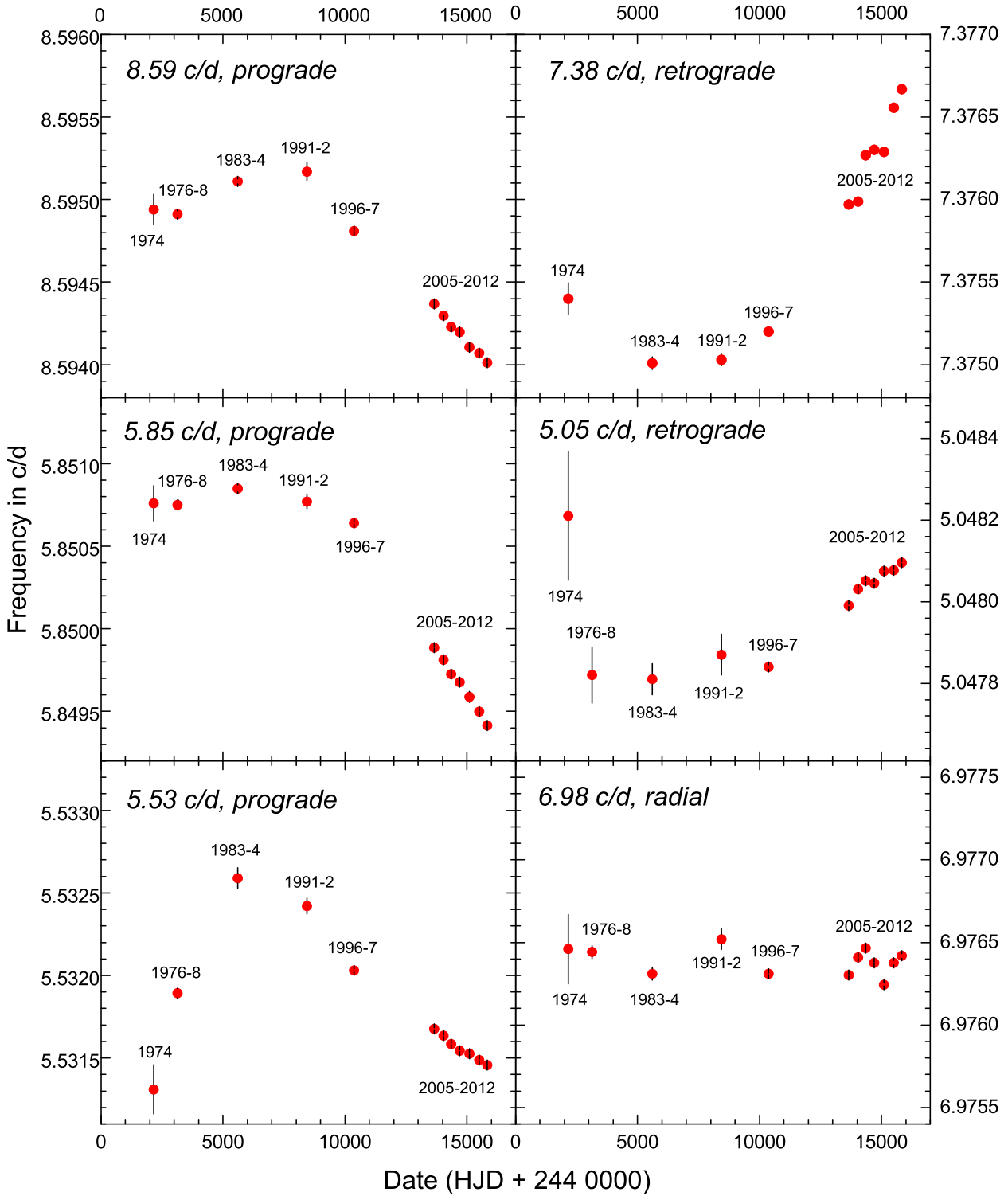}
\caption{Frequency variations of the six dominant modes from 1974 to 2012. The errors bars shown are usually the size of the symbols. The diagram shows the systematic difference between prograde and retrograde modes. Furthermore, around $\sim$1986 the sign of the frequency changes reversed systematically.}
\end{figure*}

We determined the values of the phase shifts and amplitude ratios for each observing season from 2005 to 2012. The annual values are in excellent agreement with each other. The results and their uncertainties are listed in Table 2.

\begin{table*}
\caption{Frequency values changing over four decades}
\begin{tabular}{ccrrrrrr}
\hline
\noalign{\smallskip}
Group & Average date &\multicolumn{6}{c}{Frequency}\\
(Year) &(HJD) &\multicolumn{6}{c}{cycles d$^{-1}$}\\
\noalign{\smallskip}
\hline
\noalign{\smallskip}
& & f$_1$ & f$_2$ & f$_3$ & f$_4$ & f$_5$ & f$_6$ \\
\noalign{\smallskip}
1974    &       244 2151        &       8.59494                 &       5.04821                 &       5.85076                 &       5.53131                 &       7.3754                  &       6.97646                 \\
        &               & $\pm$ 0.00009                 & $\pm$ 0.00016                 & $\pm$   0.00011                 & $\pm$ 0.00015                 & $\pm$ 0.00009                 & $\pm$   0.00021                 \\
1976 - 1978     &       244 3126        &       8.594912                        &       5.04782                 &       5.85075                 &       5.531893                        &       ...                     &       6.976442                        \\
        &               & $\pm$ 0.000019                        & $\pm$ 0.000069                        & $\pm$   0.000031                        & $\pm$ 0.000029                        &       ...                     & $\pm$   0.00004                 \\
1983 - 1984     &       244 5601        &       8.59511                 &       5.04781                 &       5.85085                 &       5.53259                 &       7.37501                 &       6.97631                 \\
        &               & $\pm$ 0.000016                        & $\pm$ 0.000038                        & $\pm$   0.000021                        & $\pm$ 0.000062                        & $\pm$   0.000038                        & $\pm$ 0.000039                        \\
1991 - 1992     &       244 8434        &       8.59517                 &       5.04787                 &       5.85077                 &       5.53242                 &       7.37503                 &       6.97652                 \\
        &               & $\pm$ 0.000055                        & $\pm$ 0.00005                 & $\pm$   0.000044                        & $\pm$ 0.000048                        & $\pm$   0.000037                        & $\pm$ 0.000063                        \\
1996 - 1997     &       245 0373        &       8.59481                 &       5.04784                 &       5.85064                 &       5.53203                 &       7.3752                  &       6.97631                 \\
        &               & $\pm$ 0.000003                        & $\pm$ 0.000005                        & $\pm$   0.000005                        & $\pm$ 0.000008                        & $\pm$   0.000004                        & $\pm$ 0.000006                        \\
2005 - 2006     &       245 3663        &       8.594369                        &       5.047991                        &       5.849885                        &       5.531677                        &       7.375969                        &       6.976303                        \\
        &               & $\pm$ 0.000003                        & $\pm$ 0.000003                        & $\pm$   0.000004                        & $\pm$ 0.000006                        & $\pm$   0.000005                        & $\pm$ 0.000005                        \\
2006 - 2007     &       245 4042        &       8.594296                        &       5.048031                        &       5.849813                        &       5.531636                        &       7.375989                        &       6.97641                 \\
        &               & $\pm$ 0.000002                        & $\pm$ 0.000002                        & $\pm$   0.000002                        & $\pm$ 0.000003                        & $\pm$   0.000004                        & $\pm$ 0.000003                        \\
2007 - 2008     &       245 4357        &       8.594227                        &       5.048051                        &       5.849725                        &       5.531585                        &       7.376268                        &       6.976467                        \\
        &               & $\pm$ 0.000002                        & $\pm$ 0.000002                        & $\pm$   0.000003                        & $\pm$ 0.000004                        & $\pm$   0.000006                        & $\pm$ 0.000004                        \\
2008 - 2009     &       245 4698        &       8.594198                        &       5.048045                        &       5.849677                        &       5.531544                        &       7.376301                        &       6.976376                        \\
        &               & $\pm$ 0.000002                        & $\pm$ 0.000003                        & $\pm$   0.000004                        & $\pm$ 0.000004                        & $\pm$   0.000007                        & $\pm$ 0.000005                        \\
2009 - 2010     &       245 5107        &       8.594107                        &       5.048075                        &       5.849587                        &       5.531526                        &       7.376288                        &       6.976244                        \\
        &               & $\pm$ 0.000002                        & $\pm$ 0.000002                        & $\pm$   0.000033                        & $\pm$ 0.000004                        & $\pm$   0.000006                        & $\pm$ 0.000009                        \\
2010 - 2011     &       245 5498        &       8.59407                 &       5.048077                        &       5.849499                        &       5.531488                        &       7.376557                        &       6.976377                        \\
        &               & $\pm$ 0.000003                        & $\pm$ 0.000002                        & $\pm$   0.000003                        & $\pm$ 0.000004                        & $\pm$   0.000006                        & $\pm$ 0.000014                        \\
2011 - 2012     &       245 5832        &       8.594011                        &       5.048096                        &       5.849415                        &       5.531456                        &       7.376668                        &       6.97642                 \\
        &               & $\pm$ 0.000003                        & $\pm$ 0.000002                        & $\pm$   0.000003                        & $\pm$ 0.000004                        & $\pm$   0.000006                        & $\pm$ 0.000014                        \\\noalign{\smallskip}
\hline
\noalign{\smallskip} \end{tabular}
\end{table*}

We can now compare the observed values with those predicted by theoretical models for different pulsation modes.
Our calculation of amplitude ratios and phase differences for the Str\"omgren $v$ and $y$ passbands relies on the formalism
presented in Daszy\'nska-Daskiewicz et al. (2003). An evolved stellar model with a mass of 2.0 solar masses was used
as input for our computation. The model parameters $\log L/L_\odot$ = 1.486, $T_{\rm eff}$ = 6845K, $\log g$ = 3.49 match
the effective temperature and luminosity determined by Schmid et al. (2014) very well
(i.e., $\log L/L_\odot$ = 1.47 $\pm$ 0.05, $T_{\rm eff}$ = 6875 $\pm$ 120K). Moreover, solar metallicity was assumed.

The comparison between the observed and the corresponding computed phase shifts and amplitude ratios is shown in Fig. 1.
Detailed inspection of the figure leads to the following conclusions: while the theoretically predicted phase differences
agree well with the observations, the amplitude ratios show a larger deviation.
This is a well known problem in $\delta$ Sct stars and is caused by the imperfect theoretical description
of the outer layers in these rather cool stars. Therefore, only the phase difference is commonly
used as a discriminator of the spherical mode degree for $\delta$ Sct stars.
As we can see from Fig. 1, the modes can be separated into three groups, each corresponding to $\ell$ = 0, 1 or 2.

We note that the derived $\ell$ values agree remarkably well with preliminary results given in Lenz, Pamyatnykh \& Breger (2010),
which used only the data from the much more limited 1996 multisite campaign. We have combined the new results with the
spectroscopically determined $m$ values. The final mode identifications are also listed in Table 2.

An interesting detail is revealed when comparing the detected azimuthal orders of the modes to their amplitude ratios.
The three modes with positive azimuthal order have lower $v/y$ amplitude ratios than their negative counterparts.
This suggests an effect of rotation on the amplitude ratios. However, this intriguing result must be regarded as somewhat
uncertain due to small-number statistics.

\section{Frequency variations: 1974 -- 2012}

Let us examine the results for the six dominant modes with known mode identifications. Fig. 2 and Table 3 show the derived
frequencies with their statistical uncertainties for each of the data groups discussed in Section 2.
We note that, except for the 1974 data, the statistical uncertainties are very small and often smaller than the size of the symbols in Fig. 2.

\begin{figure*}
\includegraphics[bb = 0 45 750 580,width=\textwidth, clip]{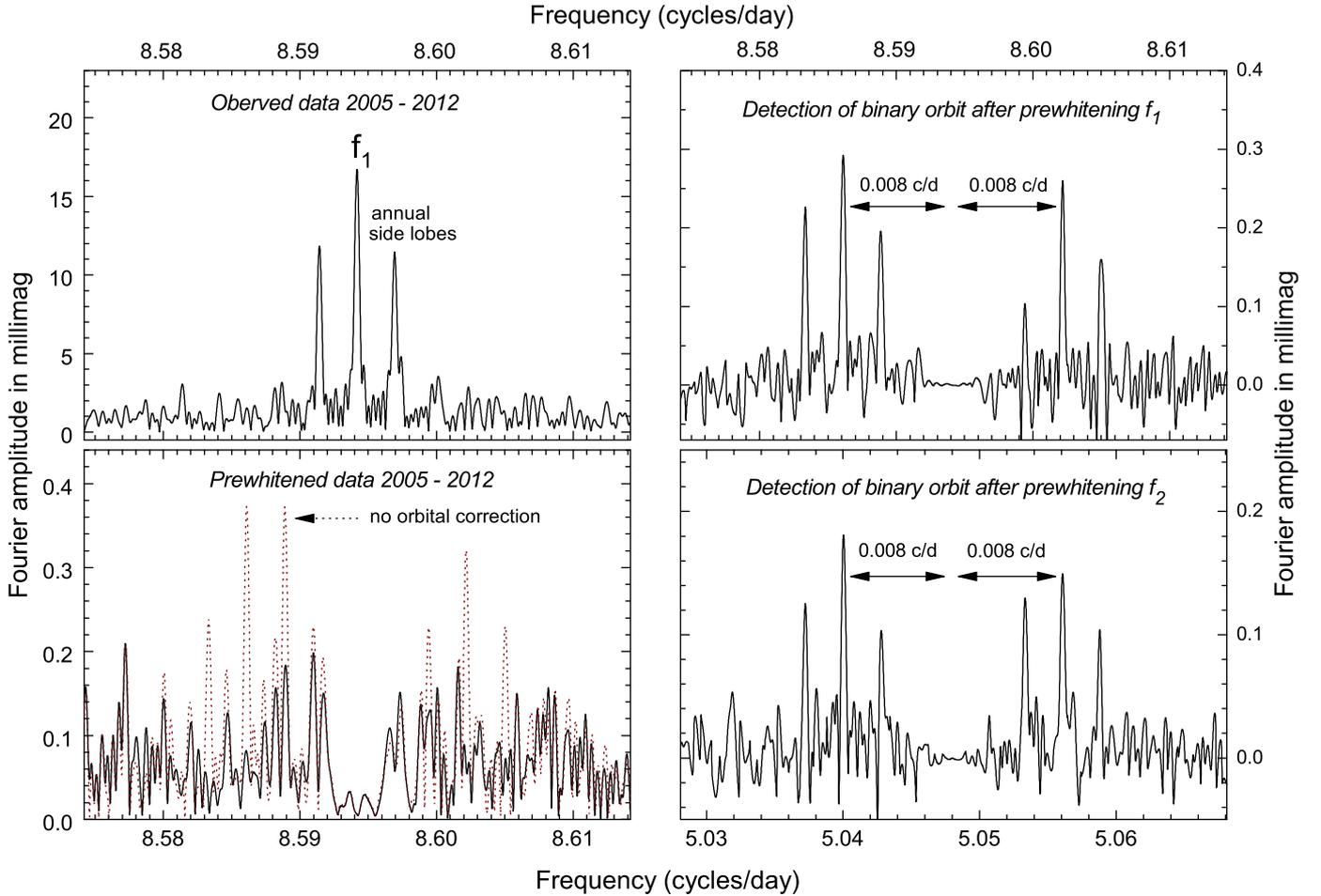}
\caption{Confirmation of the binary orbit from a comparison of the 2005 -- 2012 photometric data without and with the orbital light-time corrections. Top left panel: Fourier amplitudes of the observed data near the $f_1$ frequency. Bottom left panel: Fourier amplitudes of the residuals after prewhitening the best 64-frequency solutions. Note the changed amplitude scale. The uncorrected data (dotted curve) shows the side peaks from the 124 ($\sim$ 1/ 0.008) d orbit, while the data corrected for the spectroscopically detected binary orbit eliminates these peaks. The difference of the $f_1$ side peaks from the lower left panel is demonstrated in the top right panel. A similar analysis for $f_2$ (bottom right panel) confirms the detection.}
\end{figure*}

The pulsation frequencies of 4 CVn are systematically changing on timescales of decades. The
photometric amplitudes of the dominant pulsation modes are also variable and range from 2 to 24 millimag. 

While the timescale of the amplitude variations are also long, the amplitude variations showed no obvious patterns,
no correlations from mode to mode, and no correlations with the systematic frequency changes. Consequently, they will not be considered further in the present discussion.

The study of the frequency variations of the dominant modes show the following remarkable patterns:

\renewcommand{\theenumi}{\roman{enumi}}
\begin{enumerate}
\item The frequency of the (axisymmetric) radial mode shows very little, if any, variation from 1974 to 2012.

\item The frequencies of the nonradial modes vary in a continuous manner on a timescale of decades. From 2005 to 2012, the rates of frequency variations were essentially constant.

\item For the nonradial modes studied, we found a simultaneous sign reversal of the rates of frequency change around 1986.

\item Prograde modes  show increasing frequencies before $\sim$1986  and decreasing frequencies after $\sim$1986.

\item Retrograde modes show decreasing frequencies before $\sim$1986 and increasing frequencies after $\sim$1986.

\item The (axisymmetric) radial mode shows only small or zero frequency changes before and after $\sim$1986.

\end{enumerate}

\section{Can the binary orbit be detected photometrically?}

The long-term period changes reported in this paper were calculated with and without orbital light-time corrections. We found that the derived (long-term) frequency changes
presented in this paper are not significantly affected by the binary nature of 4 CVn. This is not surprising
since the orbital period is about 1/3 years, implying that each year covers almost three full orbits.
This means that combining data from two adjacent years covers the orbital phases in a similar way. Also, the annual observing seasons cover at least one full orbit for most years.
For the 702 nights of photometry presented in Paper I (2005 -- 2012 data),
correction of the light-time effects in the eccentric binary orbit leads to a small reduction of the residuals of the 64-frequency solution, providing support for the
spectroscopic binary detection.

In a series of recent papers (e.g., Shibahashi \& Kurtz 2012, Shibahashi et al. 2015, Murphy et al. 2016), it has been pointed out that the binary nature can also be detected from
the observed frequency modulations and phase shifts of the pulsation modes caused by the orbital motion.
This has been successfully applied to pulsators measured by the {\it Kepler} spacecraft.
If the observational time span of each data group is long enough, we expect that the Fourier spectrum of the pulsation modes
are triplets with small side lobes separated from the central component by 1/$P_{\rm orb}$. For 4 CVn we would expect side lobes separated from the intrinsic pulsation
frequencies by 0.008 cycles d$^{-1}$ and an amplitude near half a millimag for the dominant $f_1$ mode.

This does not mean that such side lobes must necessarily be detectable
in the ground-based photometry of 4 CVn: the detection of the orbital side lobes is severely affected by the
known long-term frequency changes and amplitude variability of the pulsation modes, gaps in the observational coverage,
as well as the large number of undetected pulsation modes
beyond the adopted 64 frequencies. 

In fact, the Fourier frequency spectrum of the observed 2005-2012 data does not clearly reveal the peaks,
even after comparison with the spectral window. This is illustrated for the dominant mode, $f_1$, in the top left panel of Fig. 3, where the small orbital
side lobes are hidden.

In order to search for such side lobes, we have used the data from the 702 nights of observations without applying the orbital corrections.
For each year, an optimum 64-frequency solution was calculated and prewhitened. This approach eliminated most the effects
of amplitude and long-term period variability, although the annual (as opposed to biannual) solutions may lead to some overinterpretation
of this variability. We then calculated a Fourier frequency spectrum from the residuals.

The bottom left panel of Fig. 3 shows the Fourier spectrum of the residuals near the frequency of $f_1$. Two side lobes separated from the prewhitened $f_1$ frequency by 0.008 cycles d$^{-1}$
are clearly seen. We have repeated the analysis after correcting for the known spectroscopically determined orbit. The top
right panel depicts the difference between the two Fourier spectra. The orbital side lobes are clearly seen together with their expected annual side lobes ($\pm$ 0.003 cycles d$^{-1}$).
The same result is obtained for $f_2$.

We conclude that the observed photometry confirms the spectroscopic binary detection and the orbital period near 125 (= 1/0.008) d.

\section{Astrophysical interpretations}

In this section we explore a number of possible astrophysical explanations for the observed systematic frequency changes.

\subsection{Binary system effects} 

Could both components of 4 CVn be pulsating with the systematically different changes of the pulsation frequencies corresponding to the two stellar components? This has been seen before: in the $\delta$ Sct star $\theta^2$ Tau (Breger 2005), the observed pulsation modes divide into two groups with opposite signs of period changes. For $\theta^2$ Tau the binary orbit is known. The predicted orbital light-time effects of the brighter component fit the observed period changes of the modes with frequencies smaller than 20 cycles d$^{-1}$, while the fainter component fit the changes of the frequencies higher than 20 cycles d$^{-1}$.  In principle, for 4 CVn such a scenario might even explain the general reversal of all the signs of period change around 1986, which would be caused by the geometry of the orbital motion of both components. However, in 4 CVn the known orbital period of 124 d does not fit the timescale of the observed frequency changes covering decades. Furthermore, in the binary-system model, all pulsation modes should show similar light-time shifts. The opposite is observed. Consequently, we have to reject the binary-system explanation for the observed period changes.

\subsection{Nonlinear mode interaction} 

Another theoretical explanation for large period changes is nonlinear mode interaction, for example, as proposed for XX Pyx by Handler et al. (2000). Such mode interactions  between three modes were described by Moskalik (1985). As noted by Blake et al. (2003), the large number of modes excited in most $\delta$ Sct stars makes theoretical predictions from models of such mode interactions difficult. In addition, it does not seem to naturally provide an explanation for the opposite behavior of prograde and retrograde modes. 

\subsection{Rotation} 

In the limit of slow rotation, the frequencies of stellar pulsation modes are split according to the following formula:
\begin{equation}
  f_{n\, l\, m} = f_{n\, l\, 0} + (1-C_{n\, l}) \, m \, f_{\rm rot},
  \label{om_pert}
\end{equation}
where $n$ is the radial overtone number, $l$ is the order, and $m$ is the azimuthal order of the mode; $f_{\rm rot} = \Omega/2 \pi$ is the solid-body rotation rate, and $C$ is the Coriolis coefficient, which is usually much less than 1 for p modes. If there is differential rotation, then the term for the splitting is replaced by an integral over the star of the local rotation rate, $\Omega(r,\theta)$, times the rotation kernel of a given mode. As Eq.~\ref{om_pert} illustrates, changes in $\Omega$ would produce opposite effects for prograde ($m > 0$) and retrograde ($m < 0$) modes. Thus, this interpretation could offer a natural explanation for these changes.

\begin{figure}[t]
\includegraphics[width=\columnwidth]{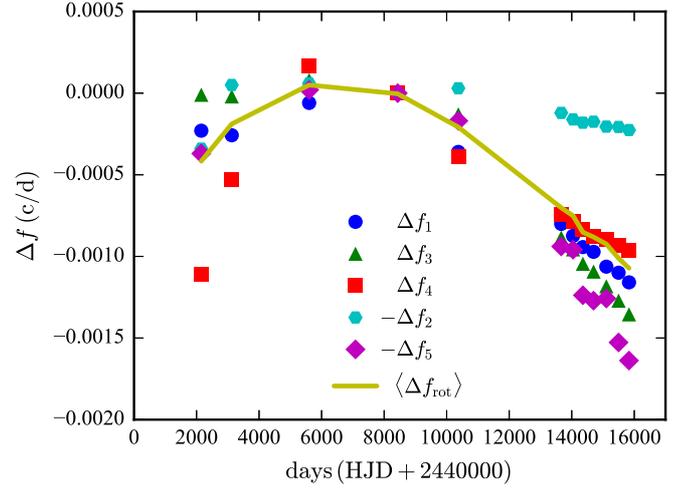}
\caption{Frequency changes of the different pulsation modes relative to their 1986 values (symbols). As indicated in the legend, we have plotted the negative of the frequency changes for the retrograde modes; the solid curve shows the average of these frequency changes.
After accounting for the sign of $m$, we see that the prograde and retrograde modes contribute morphologically similar frequency changes.}
\end{figure}

\subsubsection{Change in rotation due to a radius change} 

As the simplest possible model, let us assume that the frequency changes are driven by a change in the radius of the star. Assuming solid body rotation and conservation of angular momentum, this implies that $R^2\,f_{\rm rot}$ is constant, so a nonradial mode will have its frequency, $f_{\rm NR}$, perturbed by
\begin{equation}
  \Delta f_{\rm NR} \sim m\,\Delta f_{\rm rot} \sim -2\,m\,f_{\rm rot} \,\frac{\Delta R}{R}.
\end{equation}
The frequency of a radial mode, $f_{\rm R}$, is $\propto (G \,\langle \rho \rangle)^{1/2}$. Consequently, we find that
\begin{equation}
  \Delta f_{\rm R} = - f_{\rm R} \,\frac{3}{2} \frac{\Delta R}{R}.
\end{equation}
Thus, we expect that
\begin{equation}
  \frac{\Delta f_{\rm NR}}{\Delta f_{\rm R}} \sim \frac{f_{\rm rot}}{f_{\rm R}}.
\end{equation}

Since the measured rotational velocity for 4 CVn is greater than $100\,$km s$^{-1}$, the rotation period is of the order of one day or smaller.  Using $f_{\rm rot} \sim 1\,{\rm d}^{-1}$ and $f_{\rm R} \sim 7\,{\rm d}^{-1}$ we find that
  $\Delta f_{\rm NR}/\Delta f_{\rm R} \sim 0.1$. However, the measured values of 
  $\Delta f_{\rm NR}$ and $\Delta f_{\rm R}$ are $\sim 6\times 10^{-5}\,{\rm d}^{-1}$ and $\sim 0.5\times 10^{-5}\,{\rm d}^{-1}$, respectively, so empirically we have  
  $\Delta f_{\rm NR}/\Delta f_{\rm R} \sim 10$. Therefore, a simple radius change fails to adequately differentiate between the behaviors of radial and nonradial modes.

\subsubsection{Differential rotation} 

We now leave aside the origin of a possible change in rotation rate and simply explore what change in rotation would be required to explain the frequency changes seen in the nonradial modes.  In Fig. 4 we show the frequency changes of the different pulsation modes relative to their 1986 values (colored points). We note that we have plotted the negative of the frequency changes for the retrograde modes; after accounting for the sign of $m$, we see that the prograde and retrograde modes contribute morphologically similar frequency changes.

The solid curve in Fig. 4 shows the average of these frequency changes; it is the average change in rotation frequency implied by these shifts, assuming solid body rotation. This interpretation makes sense qualitatively, explaining the overall behavior of the prograde, retrograde, and radial modes. Unfortunately, there is a problem:
since $C_{n\,l} \ll 1$ in $\delta$~Sct models, Eq.~\ref{om_pert} implies that $\Delta f_{n\, l\, m} \approx m\, \Delta f_{\rm rot}$, that is, for a given change in rotation rate the magnitude of the frequency changes of all the non-radial modes should be nearly the same. As can be seen in Fig. 4, the frequency changes of the modes differ from each other by factors of up to two or three.

To salvage this explanation we must assume three things: (1) the star is differentially rotating, (2) different modes sample different effective rotation rates, and (3) the frequencies of the modes are more affected by rotation than is usually seen in p-mode pulsations. Given that $f_{\rm rot}/f_{\rm pulse} \ga 0.2$ is not small, the modes in this star are not in the slow rotation limit. Since Eq.~\ref{om_pert} was derived assuming slow rotation, it may not be applicable to this star. In particular, the eigenmodes we observe could be more sensitive to changes in $\Omega(r,\theta)$ than is expected in a slowly rotating star. 

This interpretation naturally explains the opposite behavior of the prograde and retrograde modes, so we find it the most compelling possibility. Even so, it may require a revision of our understanding of the effects of rotation on the pulsations of $\delta$ Sct stars in the moderate rotation regime.

As a final point, we mention that the rotation profile we hypothesize is not changing due to an external torque on the star. Rather, we imagine that there are processes, perhaps the pulsations themselves, that redistribute angular momentum within the star. Since the modes in general have different rotation kernels, this will produce different period changes for different modes.

\section{Conclusions}

We have examined over 800 nights of high-precision photometry of the evolved multiperiodic pulsator 4~CVn obtained from 1966 through 2012. Most of the data were obtained in adjacent observing seasons. This made it  possible to derive very accurate period values for a number of the excited pulsation modes and to study their systematic changes from 1974 to 2012.

Most pulsation modes of 4~CVn show systematic significant period and amplitude changes on a timescale of decades. Around 1986, a general reversal of the directions
of both the positive and negative period changes was observed in the well-studied modes. Furthermore, the period changes between the different modes
are strongly correlated, although they differ in size and sign. An important discovery is that for the modes with known ($\ell, m$) values,
there exists a correlation between the direction of the period changes and the identified azimuthal order, $m$. The associated amplitude changes
generally have similar timescales of years or decades, but show little systematic or correlated behavior from mode to mode.

To explain these observations, we have examined several potential physical explanations. Binary system effects, nonlinear mode interaction and changes in rotation due to a radius change were found to be unsatisfactory. However, a natural explanation for the opposite behavior of the prograde and retrograde modes is that their period changes are driven by a changing rotation profile. The changes in the rotation profile could in turn be driven by processes, perhaps the pulsations themselves, that redistribute angular momentum within the star. Our data show that the magnitudes of the period changes of the modes are different, which are not expected for $p$-mode pulsations in the slow rotation limit. Given that $f_{\rm rot}/f_{\rm pulse} \ga 0.2$ is not small, we see the variation in magnitude of period change as evidence that the pulsations are strongly modified by rotation.

\begin{acknowledgements}

 MB is grateful to the Austrian Fonds zur F\"orderung der wissenschaftlichen Forschung for supporting this investigation through project P 21830-N16. MHM is grateful for the support of the NSF under grant AST-1312983. AAP acknowledges partial financial support from the Polish NCN grant 2011/01/B/ST9/05448.
 \end{acknowledgements}

\end{document}